# Tuning the Fröhlich exciton-phonon scattering in monolayer MoS₂

Bastian Miller[1,3], Jessica Lindlau[2,3], Max Bommert[1,3], Andre Neumann[2,3], Hisato Yamaguchi[4],

Alexander Holleitner[1,3], Alexander Högele[2,3] and Ursula Wurstbauer[1,3,§]

[1] *Walter Schottky Institute and Physics-Department, Technical University of Munich, Am Coulombwall 4a, 85748 Garching, Germany*

[2] *Fakultät für Physik, Munich Quantum Center, and Center for NanoScience (CeNS), Ludwig-Maximilians-Universität München, Geschwister-Scholl-Platz 1, 80539 München, Germany*

[3] *Nanosystems Initiative Munich (NIM), Schellingstr. 4, 80799 München, Germany*

[4] *MPA-11 Materials Synthesis and Integrated Devices, Materials Physics and Applications Division, Los Alamos National Laboratory (LANL), Los Alamos, NM 87545, U.S.A.*

[§] *wurstbauer@wsi.tum.de*

A direct band gap[1], remarkable light-matter coupling[2] as well as strong spin-orbit[3] and Coulomb interaction[4] establish two-dimensional (2D) crystals of transition metal dichalcogenides (TMDs) as an emerging material class for fundamental studies as well as novel technological concepts. Valley selective optical excitation allows for optoelectronic applications based on the momentum of excitons[5–9]. In addition to lattice imperfections and disorder[10], scattering by phonons is a significant mechanism for valley depolarization and decoherence in TMDs[11] at elevated temperatures[8] preventing high-temperature valley polarization required for realistic applications. Thus, a detailed knowledge about strength and nature of the interaction of excitons with phonons is vital. We directly access exciton-phonon coupling in charge tunable single layer $MoS_2$ devices by polarization resolved Raman spectroscopy. We observe a strong defect mediated coupling between the long-range oscillating electric field induced by the longitudinal optical (LO) phonon in the dipolar medium and the exciton. We find that this so-called Fröhlich exciton LO-phonon interaction is suppressed by doping. This suppression correlates with a distinct increase of the degree of valley polarization of up to 20 % even at elevated temperatures of 220 K. Our result demonstrates a promising strategy to increase the degree of valley polarization towards room temperature valleytronic applications.

For both, optical as well as electronic phenomena, strength and nature of the electron-phonon interaction are crucial. Scattering with optical phonons dominates the mobility in single layer $MoS_2$ at room-temperature[12,13]. Moreover, electron-phonon coupling plays a fundamental role in the dynamics of photo excited electron hole pairs and related excitons that are bound by strong Coulomb interaction. The electron-phonon or exciton-phonon interaction is of great importance regarding fast cooling of photo-excited carriers[14–16], the homogeneous linewidth of excitonic luminescence[17–19], optical absorption spectra[20] and coherence[21]. In particular, phonon scattering limits the valley polarization and valley coherence of excitons, resulting in a breakdown of valley polarization at temperatures above ~100 K[8,11]. Exciton-phonon interaction is directly accessible by resonant Raman spectroscopy, where excitons play an important role as real intermediate states[22]. Here, we combine polarization resolved photoluminescence (PL) with resonant and non-resonant Raman spectroscopy to identify Fröhlich exciton-LO phonon interaction as a significant contribution to valley depolarization via direct exchange interaction in single layer $MoS_2$. We use field effect structures with electrolyte gates that enable a tuning of the free electron density $n_e$ by two orders of magnitude to demonstrate electronic control over the Fröhlich exciton-LO phonon scattering rate and its correlation to the degree of circularly polarization (DoP) of the PL of the A exciton.

Polarization resolved PL measurements and the resulting DoP are summarized in Fig. 1 for a large range of charge carrier densities at elevated temperature of $T = 220$ K for an excitation energy of $E_i = 1.96$ eV. The PL experiments are carried out on a 1L-$MoS_2$ field effect device utilizing an ionic liquid top gate. Fig. 1(a) shows the circularly co- and cross-polarized PL of the A exciton for applied gate voltages of -2 V and 0 V, corresponding to low and high electron densities, respectively. We estimate an increase of the electron density in the order of ~$10^{13}$ cm$^{-2}$ when increasing $V_{gate}$ by 2 V [Supplementary Information, Fig. S1]. For increasing $V_{gate}$, we observe a decrease of the PL intensity, consistent to the well-studied bleaching of the electron radiation interaction for high $n_e$ resulting mainly from Coulomb screening[23]. Fig. 1(b) shows the corresponding spectrally resolved degree of polarization calculated as DoP = (I($\sigma^+$) - I($\sigma^-$))/(I($\sigma^+$) + I($\sigma^-$)) for a series of $V_{gate}$. In Fig. 1(c) the DoP at the maxima of the PL peak is plotted as function of the gate voltage together with the DoP of the individual contributions decomposed by a line-shape analysis using Gaussian functions for the neutral ($A^0$) and the charged ($A^-$) excitons [c.f. Supplementary Information, Fig. S7]. Taking the values obtained from the total PL signal as a lower limit, we observe an increase of the DoP to up to 20 % with increasing charge carrier density $n_e$ by applying a gate voltage of $V_{gate} = 1$ V, while for depletion of the 2D system with negative $V_{gate}$, the DoP is vanishing. The values for the $A^0$ and $A^-$ contributions even reach DoP values of ~60 % and 40 %, respectively. According to literature, optically induced valley polarization is robust only for temperatures up to ~100 K[8], what is consistent with the absence of valley polarization in our measurements at $T = 220$ K for low $n_e$. It is known that resonant pumping increases the DoP[7]. In the presented experiment, however, the energy of the A exciton complex gets slightly more off-resonant for increasing $n_e$ [Fig. 1(a)]. Thus, the resonance energy cannot account for an increasing DoP, which we observed on multiple samples. We therefore investigate the exciton-phonon interaction as a possible depolarizing mechanism in dependence of $n_e$ by means of Raman spectroscopy.

The Raman active optical phonon modes in 1L-$MoS_2$ visible in backscattering configuration are the out-of-plane oscillation $A'_1$ and the in-plane mode E' that is represented by one longitudinal optical (LO) and one transverse optical (TO) phonon branch [Fig. 2(a)]. Phonons interact with electrons via the deformation potential (DP)[24]. Additionally, in polar crystals such as TMDs, longitudinal optical (LO) phonons induce a macroscopic electric field which can strongly couple to electrons or excitons via the Fröhlich interaction (FI)[25] as sketched in Fig. 2(b). In polarization resolved light scattering experiments, the observed intensity is determined by

(1)     $I \propto |\hat{e}_s \cdot \boldsymbol{\mathcal{R}} \cdot \hat{e}_i|^2$,

where $\hat{e}_i$ and $\hat{e}_s$ are the electric field vectors of the incident and the scattered light and $\mathcal{R}$ is the tensor of the scattering interaction, which in the case of DP interaction represents the symmetry of the phonon mode. For the A'$_1$ and the E' phonons, the DP Raman tensors are[26]:

$$(2) \qquad \boldsymbol{A}_{DP} = \begin{pmatrix} a & 0 & 0 \\ 0 & a & 0 \\ 0 & 0 & b \end{pmatrix}, \boldsymbol{E}_{DP}^{LO} = \begin{pmatrix} 0 & d' & 0 \\ d' & 0 & 0 \\ 0 & 0 & 0 \end{pmatrix}, \boldsymbol{E}_{DP}^{TO} = \begin{pmatrix} d & 0 & 0 \\ 0 & -d & 0 \\ 0 & 0 & 0 \end{pmatrix}.$$

For linearly polarized light, according to equations (1) and (2) the A'$_1$ mode maintains the polarization of the scattered light, whereas light scattered by the E' mode is unpolarized. In the case of circularly polarized incident light, the A'$_1$ mode maintains the polarization, whereas the E' mode turns circularly right-handed ($\sigma^+$) to circularly left-handed ($\sigma^-$) polarized light[27]. The polarization dependences of the DP tensors are confirmed in non-resonant ($E_i = 2.54$ eV) Raman measurements depicted in Fig. 2(c), where the A'$_1$ mode is co-polarized and the E' mode is cross-polarized under circularly polarized excitation. We refer to the configurations ($\hat{e}_i$, $\hat{e}_s$) = ($\sigma^+$, $\sigma^+$) and ($\sigma^+$, $\sigma^-$) as co- and cross-polarized configurations, respectively. The polar plot representation of the normalized mode intensities in Fig. 2(f) clearly shows the opposite polarization of the A'$_1$ and the E' modes under circular excitation.

In contrast to the DP Raman tensor, the tensor for scattering due to Fröhlich interaction is diagonal[28], hence, the scattering is expected to be co-polarized.

$$(3) \qquad \boldsymbol{E}_{FI}^{LO} = \begin{pmatrix} c & 0 & 0 \\ 0 & c & 0 \\ 0 & 0 & c \end{pmatrix},$$

Consequently, in TMDs the DP and FI contributions to the LO-phonon scattering are distinguishable by their contrasting polarization selection rules under excitation with circularly polarized light. Indeed, for excitation with $E_i = 1.96$ eV, which is close to the outgoing resonance with the A exciton of MoS$_2$, we observe a very strong contribution of the E' mode in the co-polarized configuration (E'$_{CO}$) in addition to a rather weak DP related cross-polarized contribution (E'$_{CROSS}$) [Fig. 2(d, g)]. Hence, the E' mode appears to be overall co-polarized. The polarization of the A'$_1$ mode remains unchanged under resonant excitation. Data is taken on a field effect structure [Supplementary Information, sample A] with polymer electrolyte top gate at a low electron density n$^0$.

The observed polarization of the E' mode strongly suggests that Fröhlich exciton-LO phonon interaction dominates the Raman scattering over the DP contribution under resonant excitation. Surprisingly, we find a strong suppression of this 'forbidden' Raman scattering for heavily electron doped MoS$_2$. Fig. 2(e) depicts resonant Raman spectra ($E_i = 1.96$ eV) for an electron density n$^{++}$ that is increased by about two orders of magnitude compared to n$^0$. We estimate the electron density $n_e$ from the energy of the A'$_1$ mode.[29] [Supplementary Information, Fig. S1]. For n$^{++}$, the intensities of the DP contributions A'$_{1CO}$ and E'$_{CROSS}$ are in the same order as for n$^0$, but E'$_{CO}$ vanishes completely such that the overall polarization dependence of the E' mode is cross-polarized [Fig. 2(h)], identical to the non-resonant spectra. In non-resonant Raman measurements, there is no change of the polarization in dependence of $n_e$ [Supplementary Information, Fig. S2]. The 'forbidden' Raman signal under resonant excitation and its suppression for large $n_e$ appears in the temperature range from 3 K to 300 K [Supplementary Information, Fig. S3].

We now turn to the discussion of the microscopic origin of E'$_{CO}$. Strong co-polarized exciton-LO-phonon scattering by FI is known from CdS, GaAs and other semiconductors[30,31], however, due to low exciton binding energies, only at low temperatures. The combined electron-phonon FI for an electron-hole pair cancels out exactly for zero phonon wave vector $q$ [32] and only the finite wave vector of the photon makes exciton-phonon Fröhlich scattering allowed in backscattering. Fig. 3(a) shows the dependence of the Fröhlich exciton-LO phonon matrix element on $qa_0$, where $a_0$ is the Bohr radius of the exciton. The interaction is strongest for $qa_0 \approx 2$. In $MoS_2$, the small exciton Bohr radius in the order of 1 nm[33] results in $qa_0 = 0.02$ for a first-order Raman process with a photon energy of $E_i = 1.96$ eV, thus, the interaction strength should be weak. However, besides this intrinsic Fröhlich exciton-LO phonon scattering, Gogolin and Rashba[34] proposed a second-order Raman process, involving Fröhlich exciton-LO phonon scattering and a second, elastic scattering process due to electron-impurity interaction, relaxing the momentum-conservation. Fig. 3(b) shows the two Feynman diagrams of the intrinsic and the impurity-assisted Fröhlich exciton-phonon Raman processes. Experimentally, the impurity-assisted second order process can be separated from the intrinsic, first order process due to interference effects as pointed out in ref. [31]. First-order scattering processes via DP or FI have the same initial and final states. Therefore, the tensors of the DP (2) and FI (3) sum up before squaring in the calculation of the scattering intensity (1):

(4)    $I \propto |\hat{e}_s \cdot (\boldsymbol{E}_{DP}^{LO} + \boldsymbol{E}_{FI}^{LO}) \cdot \hat{e}_i|^2$

In contrast, due to larger possible phonon wave vectors, the final states of the impurity-assisted second-order process are different and the scattering intensities sum up after squaring, prohibiting interference effects:

(5)    $I \propto |\hat{e}_s \cdot \boldsymbol{E}_{DP}^{LO} \cdot e_i|^2 + |\hat{e}_s \cdot \boldsymbol{E}_{FI}^{LO} \cdot \hat{e}_i|^2$

For intrinsic FI scattering, the interference in (4) leads to a variation of $I$ for different orientations of linearly polarized light with respect to the crystal axes of the sample. Fig. 3(c) shows Raman intensities for parallel polarized incident and scattered light ($\hat{e}_i \parallel \hat{e}_s$) for a whole rotation of $\hat{e}_{i(s)}$ in the plane of the $MoS_2$ crystal [spectra shown in Supplementary Fig. S4]. We compare the fitted amplitudes of the E' mode to a calculation of the expected intensities for purely intrinsic or purely impurity assisted FI scattering according to (4) and (5). For the calculation, we extract the relative amplitudes of the DP and the FI contributions from measurements with circularly polarized light. As a reference, we show the calculated and measured Raman intensities of the silicon TO mode because the Raman tensor of the Si TO mode implies an intrinsic correlation of the scattering intensity and the orientation of $\hat{e}_{i(s)}$. From the comparison of experiment and calculations, we conclude that the observed forbidden Raman scattering is consistent to an impurity assisted second-order Fröhlich exciton-LO phonon scattering process that activates scattering with large $q$ phonons. We would like to stress, that for an increase of the exciton radius $a_0$ by e.g. a factor of 10 with increasing electron density[23] and the subsequent increase of $qa_0 = 0.2$, the probability of the intrinsic process is only minor increased [Fig. 3(a)] and remains small. As $q$ is not fixed in the impurity assisted process, the $qa_0$ dependence of this interaction remains valid. Further, we exclude an externally applied off-plane electric field to be responsible for the activation of the Fröhlich interaction, because we do observe the presence and absence of the co-polarized E' phonon mode in resonance Raman scattering for samples with different intrinsic doping levels without the application of an electric field [Supplementary Information, Figs. S5, S6]. This large variation in the intrinsic doping level of exfoliated $MoS_2$ monolayers from sample to sample might explain conflicting reports in literature for as-prepared $MoS_2$ monolayers demonstrating the E' phonon being cross-polarized [27] or co-polarized [35] under resonant excitation.

The impurities involved in the scattering process can be either neutral or charged[31]. Electrostatic doping leads to screening of charged impurities as well as to a filling of (shallow) potential fluctuation, and hence to a reduction of the electron-impurity scattering cross section. As the change of the Fermi energy in our experiments is limited to ~13 meV, we expect shallow potential fluctuations induced by local strain or dielectric modifications due to interfacial imperfection or by the interaction with (charged) impurities in the substrate to be responsible for the impurity assisted Fröhlich scattering process. Additionally to the screening of impurities, the suppression of the FI scattering with increasing $n_e$ might also result from dielectric screening of the FI because the strength of the FI is inverse proportional to the dielectric constant[25]. We can exclude that a shift and broadening of the excitonic resonance and the well-studied bleaching of the absorption at the exciton resonance, resulting mainly from Coulomb screening[23], to be responsible for the complete suppression of the FI scattering intensity with increasing $n_e$. The extent to which these effects influence the scattering probability can be estimated from the comparison between the DP and the FI contributions [Supplementary Information, Fig. S5], because the electron-radiation interaction and the resonance condition is equally involved in both scattering mechanisms, as we find in temperature dependent Raman and PL measurements [Supplementary Information, Fig. S8]. The much stronger suppression of the FI contribution compared to the DP contributions therefore indicates a suppression of the scattering interaction itself. We conclude that impurity screening and/or dielectric screening of the FI are presumably the most relevant effects to account for a complete suppression of the Fröhlich scattering for high $n_e$.

The suppression of Fröhlich scattering with increasing $n_e$ coincides with an increase of the DoP of the PL from the A exciton. Excitonic intervalley scattering under electron-hole exchange interaction is forbidden by symmetry. The long-range exchange interaction between electron and hole of an exciton is an efficient exchange mechanism between s and p excitonic states in different valleys [36], whereas s and p states in the same valley do not mix. The strong long-range electric field induced by the LO phonon can efficiently brake the symmetry. The broken symmetry enables this mixing of s and p excitonic states, resulting in a loss of valley polarization via the long-range exchange interaction.

In summary, we observe an increase of the valley polarization of the A exciton with increasing electron density. In corresponding Raman measurements, we find strong polarization forbidden resonant Raman scattering from the LO phonon, which we can attribute to Fröhlich exciton-LO phonon scattering due to an impurity assisted second-order process. Electron doping suppresses this process entirely. We conclude that a reduction of the exciton-phonon scattering rate can improve the degree of valley polarization even at a temperature of 220 K. Our experiments demonstrate the relevance of Fröhlich interaction to optical processes in TMDs and uncover a promising strategy for simultaneously improving valley polarization properties and the charge carrier mobility particularly at elevated temperatures, as required for realistic (opto-) electronic device applications.

## Acknowledgements

We gratefully acknowledge financial support by the Deutsche Forschungsgemeinschaft (DFG) via excellence cluster "Nanosystems Initiative Munich", the European Research Council (ERC) under the ERC Grant Agreement no. 336749, the Volkswagen Foundation, DFG projects WU 637/4- 1 and HO 3324/9-1, the Center for NanoScience (CeNS) and LMUinnovativ.

## Author contributions

B.M., J.L. and A.N. performed the measurements. B.M., M.B, J.L., A.N. and H.Y. prepared the samples.  B.M., J.L., A.Hoe and U.W. conceived the experiment. B.M., J.L., A.Hoe, A.Hol. and U.W. analyzed the data. B.M. and U.W. prepared the figures and wrote the manuscript. All authors discussed the results and commented on the manuscript.

## Competing interests

The authors declare no competing financial interests.

**Figure 1: Valley polarization in dependence of the electron density.** (a) Circularly polarized PL spectra for $\sigma^+$ excitation and $\sigma^+$ (co-polarized) and $\sigma^-$ (cross-polarized) detection measured with $E_i = 1.96$ eV and at $T = 220$ K for two different gate voltages of a 1L-MoS$_2$ device with ionic liquid top gate. Spectra are normalized to the maximum of the respective co-polarized spectrum. The absolute intensity of the spectra for -2 V is a factor of 10 higher than for the spectra taken for 0 V. (b) Spectrally resolved degree of polarization for a series of gate voltages. Negative (positive) gate voltages correspond to electron depletion (accumulation). (c) Degree of polarization as a function of the applied top gate voltage of the total PL signal as shown in (b) evaluated at the PL peak maxima, and of the individual contributions of the neutral (A$^0$) and charged (A$^-$) exciton obtained from peak fits.

**Figure 2: Polarization of phonon modes in dependence of the charge carrier density.** (a) Raman active optical phonons in $MoS_2$: the in-plane, polar E' mode and the out-of-plane, homopolar A'$_1$ mode. (b) Illustration of the movement of the atoms for the LO phonon mode. The resulting electric field is indicated with red arrows. The interaction strength between the macroscopic electric field and an exciton depends on the ratio between the exciton radius and the phonon wave vector. (c-e) Polarization resolved Raman spectra for circularly polarized light from a 1L- $MoS_2$ flake in a field effect device with polymer electrolyte gate at $T = 300$ K. The filled curves are Lorentzian fits to the data. (c) Non-resonant excitation and low charge carrier density $n^0$ ($V_{TG}$ = -0.5 V, $V_{BG}$ = -40 V). (d) Resonant excitation and low charge carrier density $n^0$. (e) Resonant excitation and high charge carrier density $n^{++}$ ($V_{TG}$ = 0 V, $V_{BG}$ = 0 V). Asterisks mark additional Raman signatures that are visible under resonant excitation and that are subject to discussion in literature. (f-h) Polar-plots of the normalized amplitude of the fitted peaks shown in the panel above the respective plot versus the rotation of the quarter wave plate. The black arrows mark 0°; 0° and 90° correspond to the ($\sigma^+$, $\sigma^+$) and ($\sigma^+$, $\sigma^-$) configurations, respectively.

**Figure 3: Impurity assisted Fröhlich scattering.** (a) Plot of the matrix element of the exciton-phonon Fröhlich interaction in dependence of the product of the phonon wave vector $q$ and the exciton radius $a_0$. See Supplementary Information for details. For the intrinsic first-order process $qa_0 \approx 0.02$, while in the impurity assisted process $q$ can take arbitrary values. The $qa_0$ dependence of $|H_{\text{Fl}}|^2$ is qualitatively independent of $a_0$. (b) Feynman diagrams for the scattering of a photon with frequency $\omega$ and momentum $k$ from initial state i to final state s by emitting a phonon with frequency $\Omega$ and momentum $q$. Upper panel: intrinsic first-order Raman process. Lower panel: second-order process involving elastic scattering with an impurity. $H_{\text{eR}}$ denotes the electron-radiation interaction; $H_{\text{eL}}$ is the electron lattice interaction, which can be either DP or FI. $H_{\text{e-i}}$ represents the electron-impurity interaction for the elastic scattering with momentum transfer $q$'. We show only one permutation of the interactions. (c) Resonant Raman intensities ($E_i = 1.96$ eV) for linear parallel polarization ($\hat{e}_i = \hat{e}_s$) for one whole rotation of the angle $\theta$ between $\hat{e}_{i,s}$ and the crystal axes. Lower panel: amplitude of the E' mode (yellow triangles). Fitted amplitudes are plotted as scatters. The line plots show the calculated polarization dependences of the intrinsic and the impurity-assisted exciton-LO phonon scattering processes (solid and dashed lines, respectively). Upper panel: amplitude of the TO mode of the silicon substrate used as a reference signal (Scatters: measured data, Line plot: simulated data). Spectra are shown in the Supplementary Information.

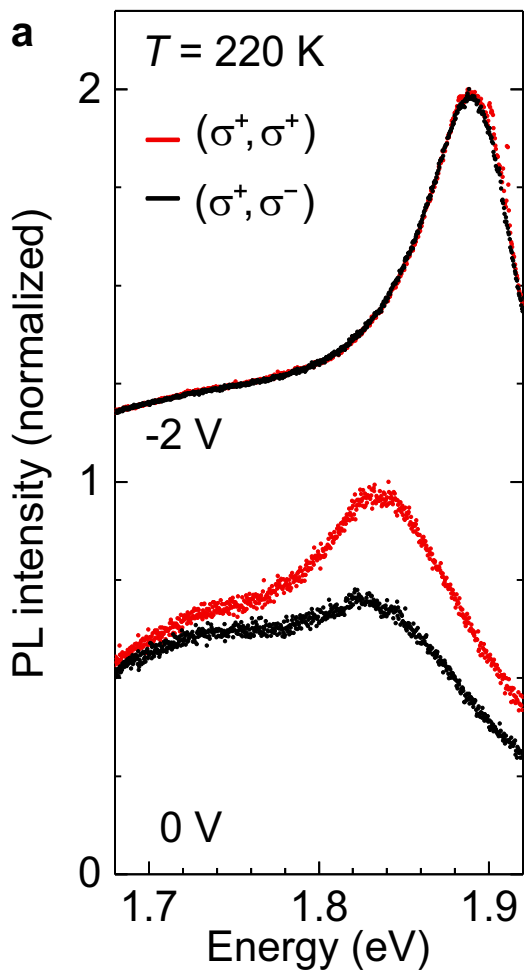

**a**

$T = 220$ K

(σ⁺,σ⁺) — red

(σ⁺,σ⁻) — black

PL intensity (normalized)

-2 V

0 V

1.7   1.8   1.9
Energy (eV)

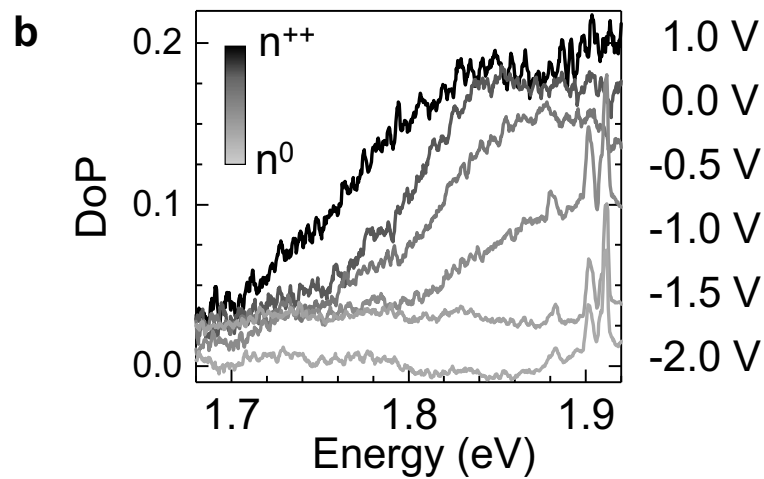

**b**

DoP

n⁺⁺
n⁰

1.0 V
0.0 V
-0.5 V
-1.0 V
-1.5 V
-2.0 V

1.7   1.8   1.9
Energy (eV)

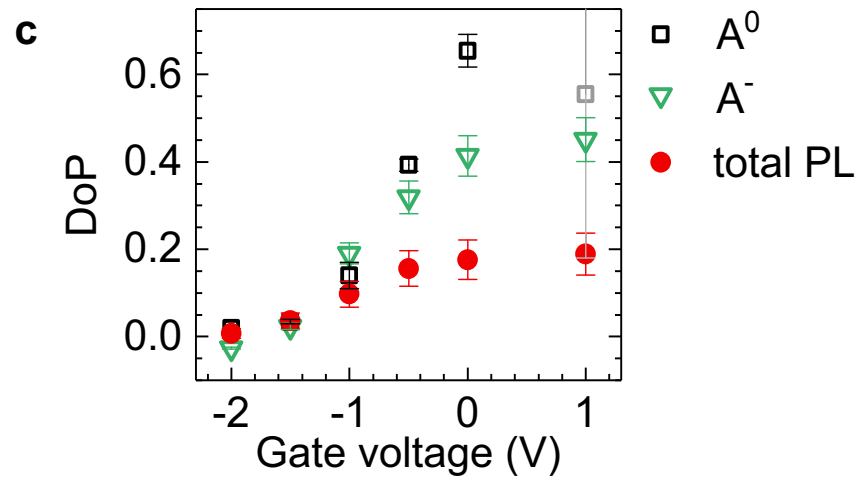

**c**

DoP

□ $A^0$
▽ $A^-$
● total PL

-2   -1   0   1
Gate voltage (V)

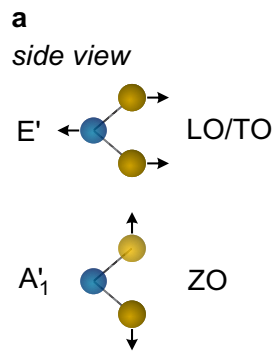

**a**

*side view*

E'  LO/TO

A'₁  ZO

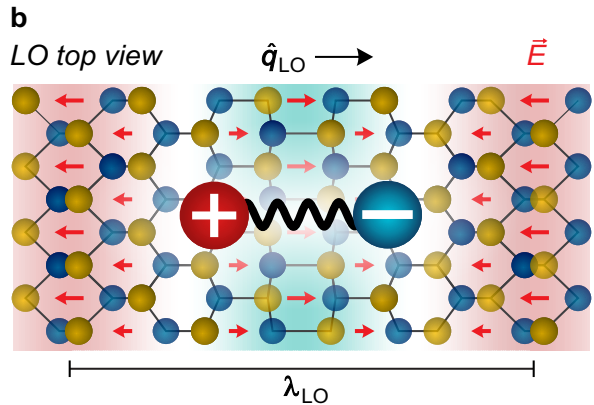

**b**

*LO top view*  $\hat{q}_{LO}$ →  $\vec{E}$

$\lambda_{LO}$

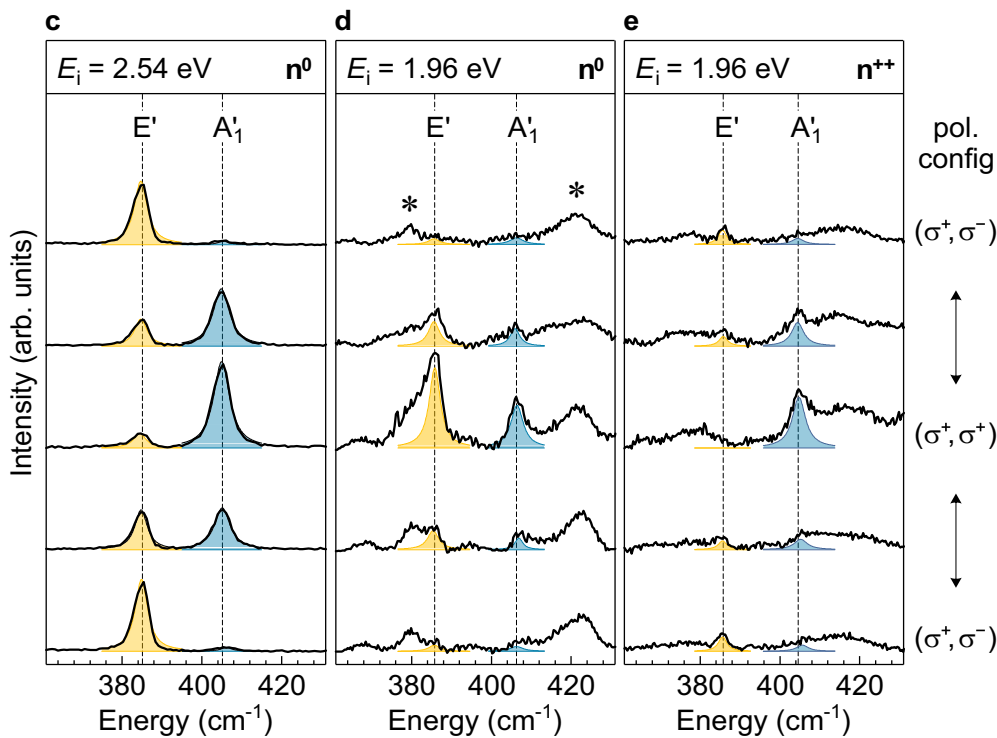

**c** $E_i$ = 2.54 eV  $n^0$

**d** $E_i$ = 1.96 eV  $n^0$

**e** $E_i$ = 1.96 eV  $n^{++}$

E'  A'₁

Intensity (arb. units)

Energy (cm⁻¹)

pol. config

$(\sigma^+, \sigma^-)$

$(\sigma^+, \sigma^+)$

$(\sigma^+, \sigma^-)$

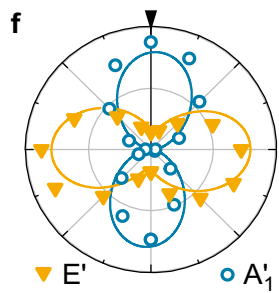

**f**

▼ E'  ○ A'₁

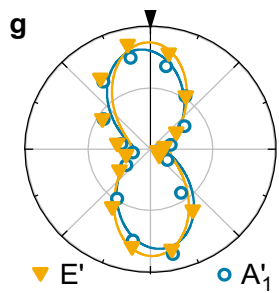

**g**

▼ E'  ○ A'₁

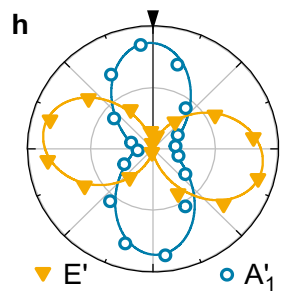

**h**

▼ E'  ○ A'₁

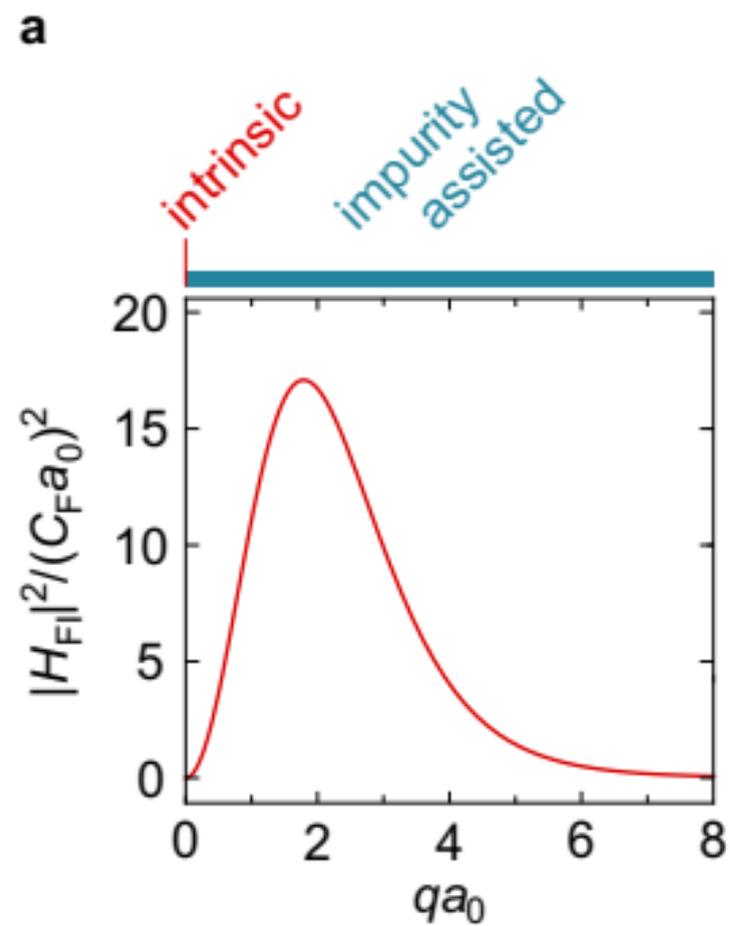

**a**

intrinsic    impurity assisted

$|H_{FI}|^2/(C_F a_0)^2$

$qa_0$

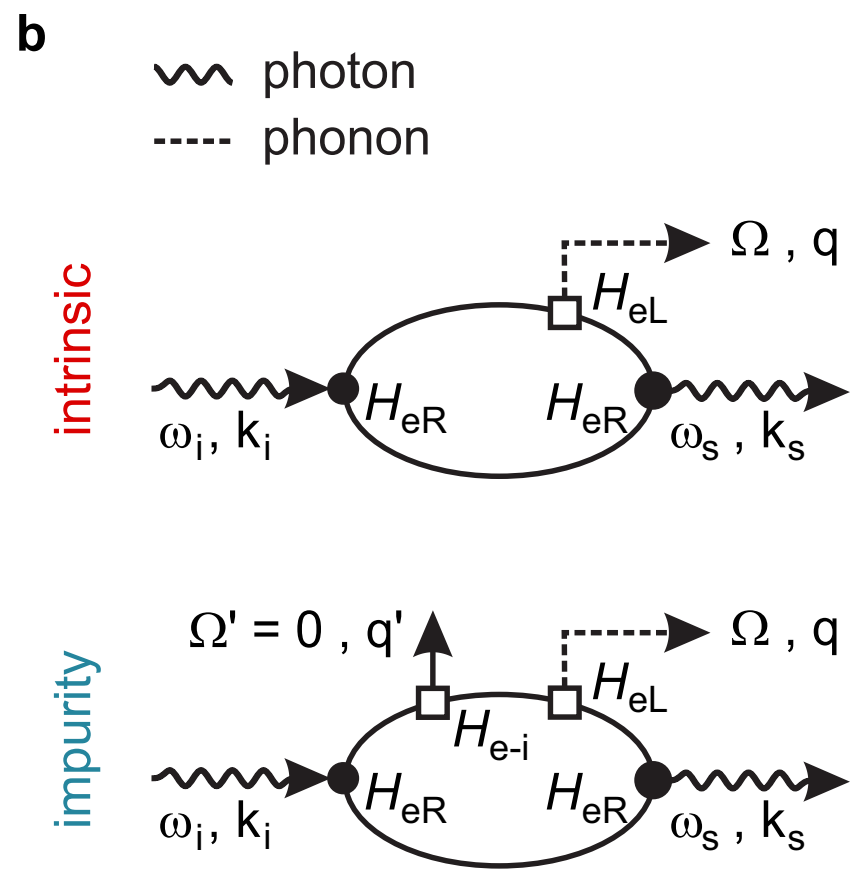

**b**

〜〜〜 photon
– – – phonon

intrinsic

$\Omega$ , q

$H_{eL}$

$\omega_i$, $k_i$    $H_{eR}$    $H_{eR}$    $\omega_s$ , $k_s$

impurity

$\Omega' = 0$ , q'    $\Omega$ , q

$H_{e\text{-}i}$    $H_{eL}$

$\omega_i$, $k_i$    $H_{eR}$    $H_{eR}$    $\omega_s$ , $k_s$

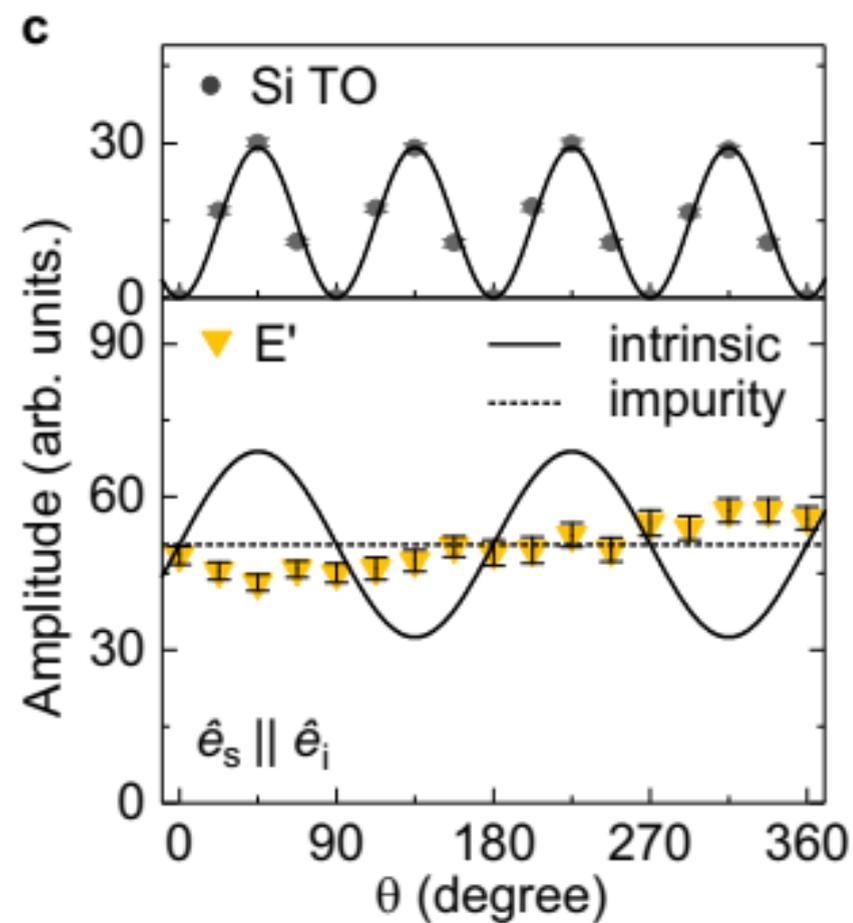

**c**

● Si TO

▼ E'    —— intrinsic
          ······ impurity

Amplitude (arb. units.)

$\hat{e}_s \parallel \hat{e}_i$

$\theta$ (degree)

Bastian Miller[1,3], Jessica Lindlau[2,3], Max Bommert[1,3], Andre Neumann[2,3], Hisato Yamaguchi[4], Alexander Holleitner[1,3], Alexander Högele[2,3] and Ursula Wurstbauer[1,3]

1. *Walter Schottky Institute and Physics-Department, Technical University of Munich, Am Coulombwall 4a, 85748 Garching, Germany*

2. *Fakultät für Physik, Munich Quantum Center, and Center for NanoScience (CeNS), Ludwig-Maximilians-Universität MÃijnchen, Geschwister-Scholl-Platz 1, 80539 München, Germany*

3. *Nanosystems Initiative Munich (NIM), Schellingstr. 4, 80799 München, Germany*

4. *MPA-11 Materials Synthesis and Integrated Devices, Materials Physics and Applications Division, Los Alamos National Laboratory (LANL), Los Alamos, NM 87545, U.S.A.*

**Contents**

## S1 Field effect devices with electrolyte top gate

Figure S1(a) shows a scheme of a field effect device we use for Raman and PL spectroscopy in dependence of the free electron density. We use a PDMS stamping technique [1] to transfer monolayer $MoS_2$ flakes onto silicon substrates with a 300 nm thick $SiO_2$ layer as dielectric. Contacts to the $MoS_2$ flake and for the electrolyte top gate are fabricated by standard optical lithography and e-beam evaporation of 5 nm Ti and 30 nm Au. As an electrolyte top gate we use either a solid polymer electrolyte (PE) of poly-(ethylene oxide) and $CsClO_4$ or the ionic liquid (IL) Diethyl-methyl-(2-methoxyethyl)-ammonium-bis-(trifluormethylsulfonyl)-imid (Sigma Aldrich). In both cases, the gating principle relies on the separation of ions in the electrolyte and the formation of an electronic double layer between the ions of the electrolyte and the two-dimensional material. The data shown in Fig. 2 of the manuscript origin from a field effect device with PE top gate [sample A]. The capacity of PE top gates is in the order of $\mu$F, two orders of magnitude higher than the capacity of the silicon back gate [2, 3]. For monolayer $MoS_2$, Chakraborty *et al.* reported a modification of the charge carrier density of $\sim 10^{13}$ cm$^{-2}$ for an applied gate voltage of $V_{TG} = 1$ V in a PE field effect device [3]. Further, the authors demonstrated that the energy of the $A_1'$ phonon mode is sensitive to the electron density and they correlated the energy shift of the $A_1'$ mode to a change of the electron density. As a determination of the absolute charge carrier density in the $MoS_2$ flake is beyond the scope of our devices, we use ref. 3 to estimate the change of the charge carrier density from the energy shift of the $A_1'$ mode measured by Raman spectroscopy. Fig. S1(c) shows non-resonant Raman spectra in the circular co-polarized configuration for two gate voltages $V_{TG} = -0.5$ V (corresponding to n$^0$ of sample A with PE gate) and $V_{TG} = 0$ V (corresponding to n$^{++}$ of sample A with PE gate). Lorentzian fits to the data reveal energies of 405.5 cm$^{-1}$ and 403.3 cm$^{-1}$ for the $A_1'$ mode, respectively. The shift of 2.2 cm$^{-1}$ corresponds to a change of the electron density of $\sim 10^{13}$ cm$^{-2}$ according to ref. [3]. For n$^0$, we assume an electron density in the order of $\sim 10^{11}$ cm$^{-2}$, what is supported by the comparison of our PL spectra to the spectra shown in ref. [4]. Overall, the estimation results in a ratio of n$^{++}$/n$^0 \sim 100$. In order to avoid asymmetric electric fields in the $MoS_2$ flake, we simultaneously use the electrolyte top gate and the silicon back gate with a ratio of $V_{BG}/V_{TG} = 80$. We would like to note, however, that an asymmetric field does not affect the conclusions drawn in the manuscript [*cf.* Fig. S6].

One peculiarity of the PE is its helical crystallization. Therefore, it can cause changes in the degree of circular polarization by turning circular polarized light into linear polarized light and by causing depolarization. For this reason, the measurements on the PE gate were conducted in the configuration shown in Fig S1(b), which makes it possible to monitor the degree of circular polarization during the measurements individually for each device since the change in the degree of circular polarization caused by the PE gate changes from sample to sample. In particular, it is possible to measure the effect of the PE gate by comparing the reflected laser light from two spots on the bare silicon substrate and on silicon substrate covered with PE. The measurements are shown in the first two rows of Fig. S1(e). The polarization dependence of the intensities of the Raman signal can be corrected for the effect of the PE gate by deconvolution. The polarization dependence of the fitted amplitudes of the $A_1'$ and the $E'$ modes is shown in Fig. S1(e) before and after the deconvolution of the effect of the PE gate. The deconvoluted data corresponds to the data shown in Fig. 2(f)-(g) of the manuscript.

The data shown in Fig. 1 of the manuscript origins from a field effect device with IL top gate [sample B]. ILs are widely used to modulate the carrier density in two-dimensional materials and they provide capacities comparable to those of PE gates [5, 6]. Fig. S1(d) shows non-resonant Raman spectra from a field effect device with IL top gate [sample B] in the circular co-polarized configuration for two gate voltages $V_{TG} = -2\,\mathrm{V}$ and $V_{TG} = 1\,\mathrm{V}$ (corresponding to $n^0$ and $n^{++}$ of sample B with IL gate, *cf.* Fig. 1 of the manuscript). Lorentzian fits to the data reveal energies of $403.8\,\mathrm{cm}^{-1}$ and $400.2\,\mathrm{cm}^{-1}$ for the $A_1'$ mode, respectively. The shift of $3.6\,\mathrm{cm}^{-1}$ verifies an equally high capacity of the IL top gate.

## S2 Matrix element of the Fröhlich exciton-phonon interaction

The Fröhlich exciton-phonon interaction for an exciton is derived by combining the electron-phonon Fröhlich interaction of an electron and a hole [7, 8]:

$$H_{\mathrm{FI}} = \frac{C_{\mathrm{F}}}{q} \left( \frac{1}{[1 + (p_{\mathrm{h}} a_0 q/2)^2]^2} - \frac{1}{[1 + (p_{\mathrm{e}} a_0 q/2)^2]^2} \right)$$

$$C_{\mathrm{F}} = e \left[ \frac{2\pi\hbar\omega_{\mathrm{LO}}}{NV} \left( \varepsilon_\infty^{-1} - \varepsilon_0^{-1} \right) \right]^{1/2}, \quad p_{\mathrm{e}} = \frac{m_{\mathrm{e}}}{m_{\mathrm{e}} + m_{\mathrm{h}}} \quad \text{and} \quad p_{\mathrm{h}} \frac{m_{\mathrm{h}}}{m_{\mathrm{e}} + m_{\mathrm{h}}}$$

For the electron and hole masses we use $m_{\mathrm{e}} = 0.46$ and $m_{\mathrm{h}} = 0.54$ [9]. For the exciton radius $a_0$ we assume $a_0 = 1\,\mathrm{nm}$ [10]. $C_{\mathrm{F}}$ contains the frequency of the LO phonon $\omega_{\mathrm{LO}}$, the volume $V$, the number of unit cells per unit volume $N$ and the high and low frequency dielectric constants $\varepsilon_\infty$ and $\varepsilon_0$. For the plot shown in Fig. 3(a) of the manuscript we assume $C_{\mathrm{F}}$ to be a constant and set it arbitrarily to $10^{-4}$.

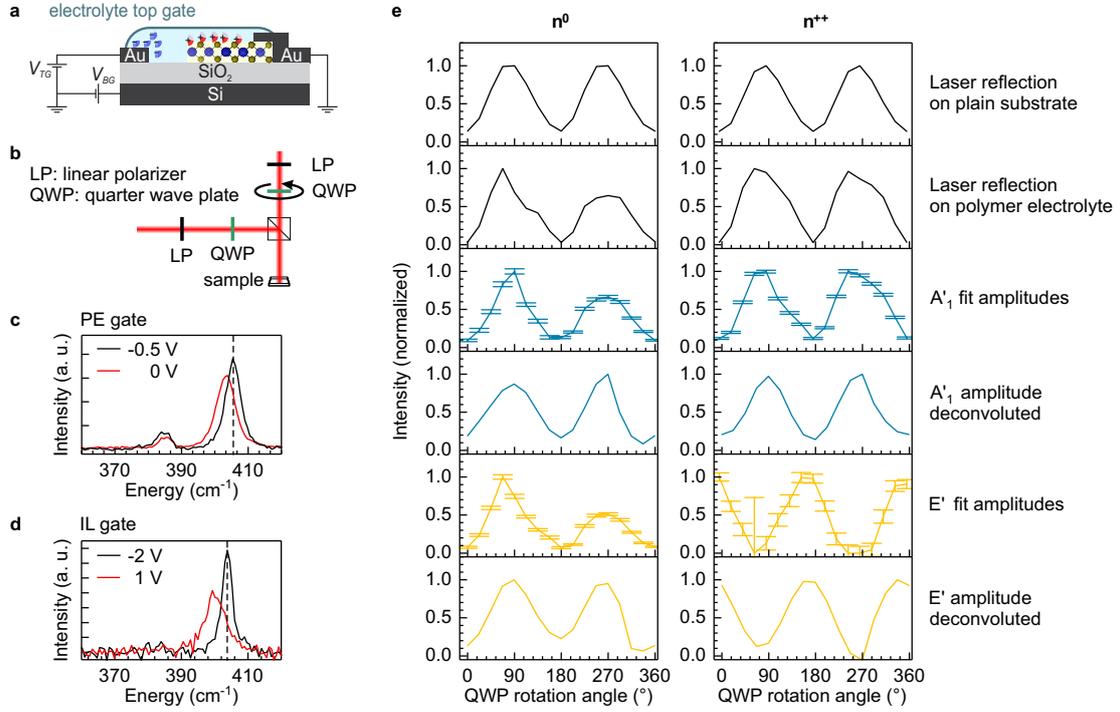

**Figure S1 — Characterization of field effect devices.** (a) Scheme of a field effect device used for optical measurements in dependence of the free electron density. (b) Scheme for polarization resolved measurements on devices with PE gate. (c) Non-resonant Raman spectra ($E_i = 2.54\,\text{eV}$) in circular co-polarized configuration for two top gate voltages of a device with PE gate [sample A]. (d) Non-resonant Raman spectra ($E_i = 2.54\,\text{eV}$) in circular co-polarized configuration for two top gate voltages of a device with IL gate [sample B]. (e) Deconvolution of the polarizing effect of the PE top gate [sample A]. The left panel shows data for low electron density $n^0$, the right panel shows data for high electron density $n^{++}$. The error bars in the plots of the fitted amplitudes represent the fit errors.

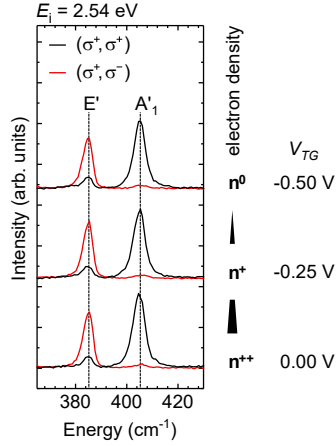

**Figure S2 — Gate dependence of non-resonant Raman spectra.** Non-resonant Raman spectra ($E_i = 2.54\,\text{eV}$) in circular co-polarized (black) and cross-polarized (red) configurations for electron densities $n^0$, $n^+$ and $n^{++}$. Data taken on sample A with PE top gate.

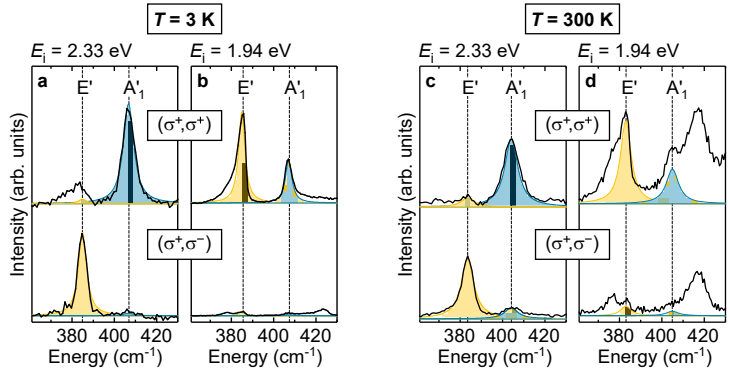

**Figure S3 — Temperature dependence of Raman spectra of CVD grown MoS₂.** Circular co-polarized ($\sigma^+,\sigma^+$) and cross-polarized ($\sigma^+,\sigma^-$) Raman spectra of a CVD grown monolayer MoS₂ flake on a SiO₂/Si substrate. Black lines represent measured spectra; filled curves are Lorentzian fits to the data. (a) Low temperature and non-resonant excitation, (b) low temperature and resonant excitation, (c) room temperature and non-resonant excitation and (d) room temperature and resonant excitation. The polarization dependence is qualitatively the same for both temperatures and it matches the data from exfoliated flakes for the case of low electron densities shown in the main part of the manuscript. We note that exfoliated MoS₂ monolayers can intrinsically be in the low or high electron density regime. This large variation in the intrinsic charge carrier density from sample to sample might explain conflicting reports in literature for pristine MoS₂ monolayers demonstrating the E' phonon being cross-polarized [11] or co-polarized [12] under resonant excitation.

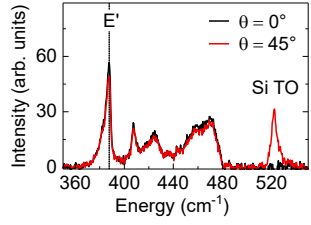

**Figure S4 — Linear polarized Raman spectra.** Resonant Raman spectra ($E_i = 1.96$ eV) for two different angles $\theta$ between the crystal axes of the sample and the polarization direction of the linear polarized incident and scattered light. Incident and scattered light are linear parallel polarized. Fitted amplitudes are shown in Fig. 3(c) of the manuscript.

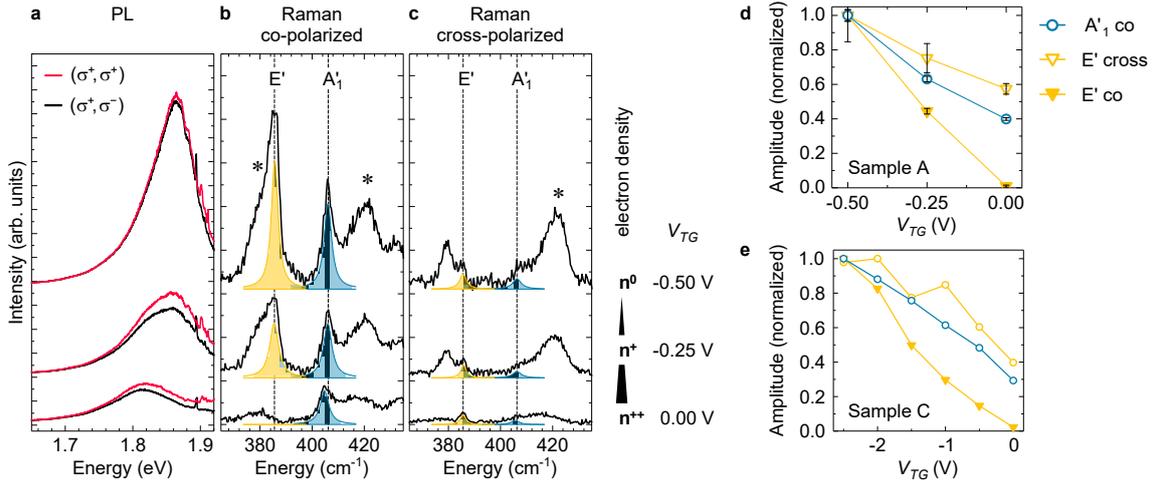

**Figure S5 — Gate dependence of PL and resonant Raman spectra.** PL and resonant Raman spectra taken at $T = 300$ K with a laser energy of ($E_i = 1.96$ eV) for three different electron densities $n^0$, $n^+$ and $n^{++}$. (a) PL spectra in the circular co- (red) and cross-polarized configuration. (b) Circular co-polarized and (c) cross-polarized resonant Raman spectra. Filled curves show fitted Lorentzian peaks for the $A_1'$ and $E'$ modes. (d) Fitted amplitudes of the Raman and PL intensities in dependence of the top gate voltage. DP contributions ($A_{1CO}'$ and $E_{CROSS}'$) are plotted as open circles, the FI contribution ($E_{CO}'$) as filled triangles. Data is shown for two samples with PE top gate. Data shown in panels (a-d) as well as data in Figs. S1(c,e), S2, S6 and Fig. 2 of the manuscript is taken on sample A.

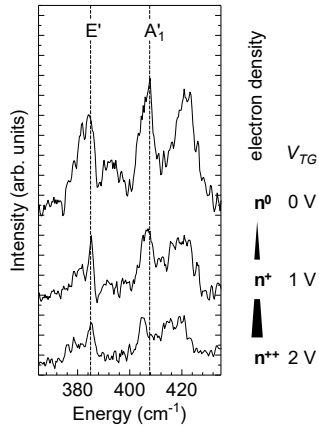

**Figure S6 — Gate dependence of Raman spectra for positive gate voltages.** Resonant Raman spectra taken at $T = 300\,\text{K}$ with a laser energy of $(E_i = 1.96\,\text{eV})$ for three different positive gate voltages on a sample with IL top gate. For $V_{\text{TG}} = 0\,\text{V}$ we observe strong scattering from the LO mode E′, while the scattering rate decreases for increasing positive gate voltages. In contrast, for sample A we observe that the LO scattering is suppressed for $V_{\text{TG}} = 0\,\text{V}$ and is strongly enhanced for negative voltages [*cf.* Fig. S5]. The difference between the two samples can be explained by different intrinsic doping levels due to the exfoliation process. Due to the contrasting behavior of the two samples when increasing the absolute of the gate voltage, we exclude symmetry breaking by the electric field of the gate to be responsible for the observation of strong co-polarized LO scattering. This conclusion is consistent to the observation of strong co-polarized LO scattering in samples without any top gate as shown in Fig. S3.

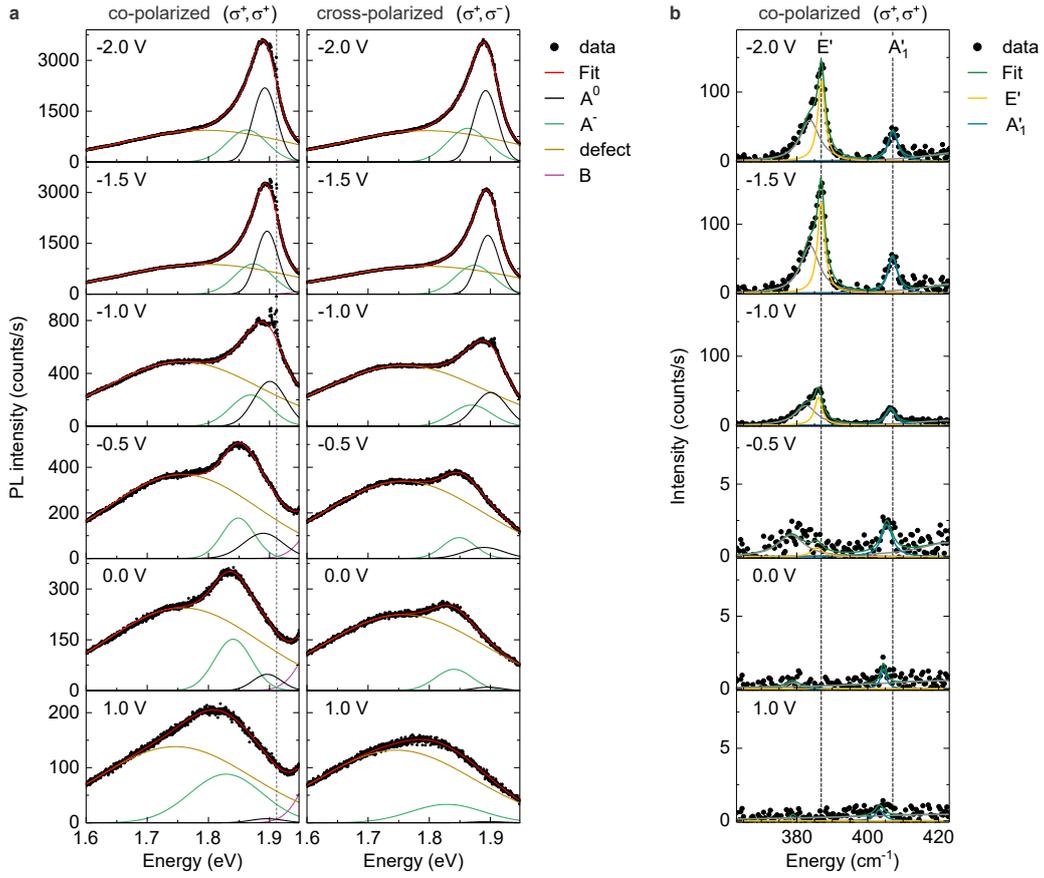

**Figure S7 — Line-shape analysis of PL and Raman spectra.** (a) Circular co- and cross-polarized PL spectra (left and right panel of (a), respectively) in dependence of the applied top gate voltage [data also shown in Fig. 1 of the manuscript]. The plots show the original data (black scatter) together with a multi-peak fit consisting of four Gaussian peaks that represent the PL of the neutral and charged A exciton ($A^0$ and $A^-$), the B exciton and a defect peak. For increasing gate voltage from $-2$ V to $1$ V we observe a bleaching of both the neutral and the charged exciton emission. The $A^0$ peak broadens from $\approx 45$ meV to $\approx 70$ meV and its energy slightly blue shifts by $\approx 4$ meV $\pm$ 10 meV. The $A^-$ peak redshifts by 35 meV, corresponding to half of its full width at half maximum of $\approx 70$ meV. The trends are consistent to existing literature for $MoS_2$ [4] and other TMDs [13]. The dashed line indicates the energy of light scattered by the $E'$ phonon and shows that the resonance condition for the light scattering is satisfied for all gate voltages. (b) Circular co-polarized Raman spectra corresponding to the spectra PL spectra shown in (a). Measured data is shown as black scatters. Solid lines are Lorentzian peaks fitted to the data. Grey lines are peaks of resonant Raman modes that are discussed in literature. In circular co-polarized configuration the $A'_1$ mode is a contribution due to the deformation potential, whereas we ascribe the $E'$ mode contribution [yellow line] to the Fröhlich interaction. For increasing gate voltages we observe decreasing intensities for both contributions, however the intensity of the $E'$ mode decreases faster than that of the $A'_1$ mode [intensity ratio plotted in Fig. S9(c)].

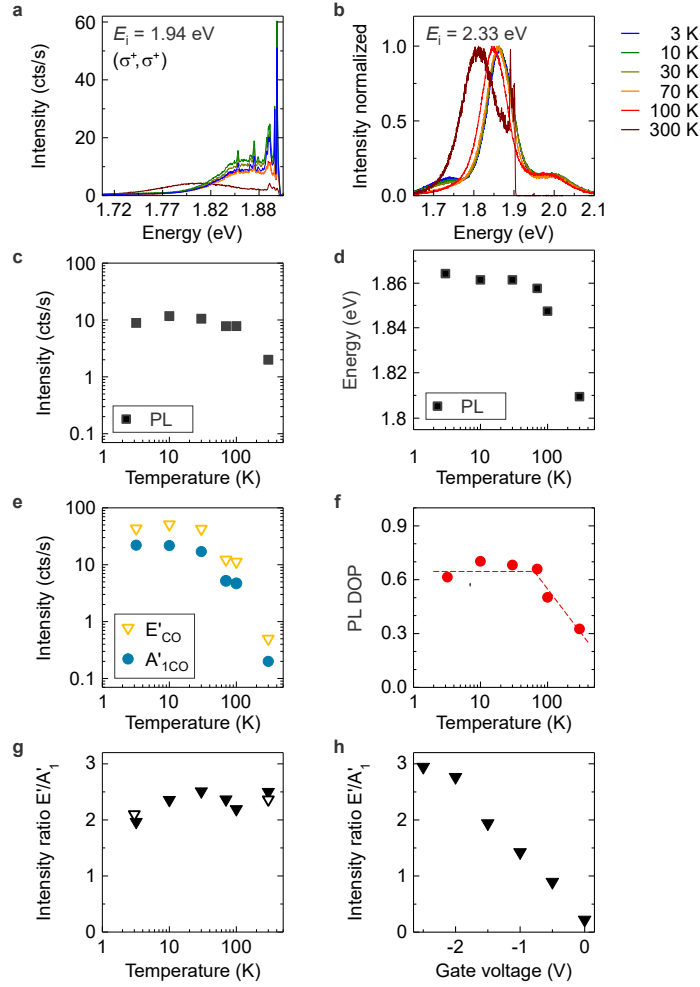

**Figure S8** — **Temperature dependence of PL and resonant Raman spectra.** Temperature dependent spectra of CVD grown monolayer MoS$_2$ [*cf.* Fig. S3]. (a) Circular co-polarized spectra for resonant excitation at $E_i = 1.94$ eV (b) Normalized spectra for excitation at $E_i = 2.33$ eV. The spectrum for 300 K is recorded with $E_i = 1.94$ eV. (c) Logarithmic plot of the PL intensity in dependence of the temperature extracted from the resonant spectra shown in (a). (d) Center energy of the PL peak in dependence of the temperature extracted from the spectra shown in (b). (e) Temperature dependence of the intensity of the co-polarized Raman contributions E$'_{CO}$ and A$'_{1CO}$ extracted from the resonant spectra in (a). The intensity of both modes decreases with increasing temperature. This effect is consistent to the bleaching and the shift of the excitonic resonance [shown in (c) and (d), respectively]. (f) Temperature dependence of the degree of polarization (DOP) of the PL spectra shown in (a). The DOP decreases monotonously with increasing temperature. (g) Ratio of the intensities of E$'_{CO}$ and A$'_{1CO}$ plotted in (e). The intensity ratio is constant over the whole temperature range, demonstrating that the temperature dependent bleaching and shift of the resonance equally affects the scattering probabilities of the DP contribution A$'_{1CO}$ and the Fröhlich contribution E$'_{CO}$. (h) Gate voltage dependence of the ratio of the intensities of E$'_{CO}$ and A$'_{1CO}$ of sample C with PE top gate [*cf.* Fig. S5(e)]. The data shows that the intensity of E$'_{CO}$ decreases much faster than the intensity of A$'_{1CO}$ with increasing electron density. From the temperature dependence shown in (g) we can exclude that the gate dependence of the resonance condition is responsible for the faster suppression of the E$'$ mode. Therefore, we conclude that this effect is caused by screening of the Fröhlich scattering with increasing electron density.

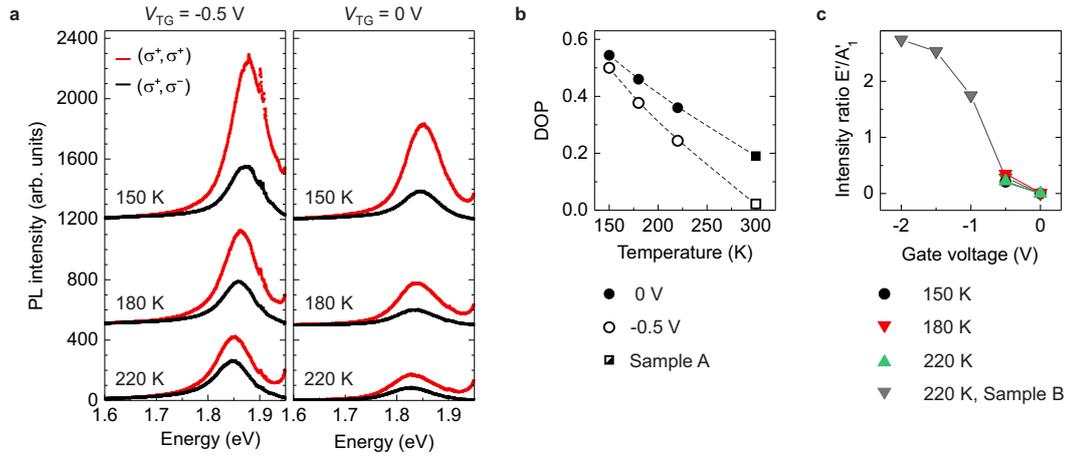

**Figure S9 — Temperature dependence of the degree of polarization.** (a) Temperature series of co- and cross-polarized PL spectra of a monolayer MoS$_2$ device with IL top gate for two different electron densities. (b) Degree of polarization (DOP) of the spectra shown in (a) evaluated at the respective maxima of the PL peak (circles). Squares represent the DOP of the PL spectra shown in Fig. S5 [sample A]. Dashed lines are guides to the eyes. (c) Intensity ratio of the E$'$ and the A$_1'$ Raman modes of circular co-polarized spectra. Grey triangles represent data taken on sample B [sample of Fig. 1 of the manuscript, spectra shown in Fig. S7(b)].